\documentclass[12pt]{article}
\usepackage{a4wide}
\usepackage{latexsym}
\usepackage{cite}
\usepackage{axodraw}

\usepackage{pslatex}
\usepackage[latin1]{inputenc}
\usepackage[T1]{fontenc}

\def\bq{\begin{eqnarray}}
\def\eq{\end{eqnarray}}
\def\l{\langle}
\def\r{\rangle} 
\def\eps{\varepsilon}

\begin{document}

\thispagestyle{empty}

\begin{flushright}
  MZ-TH/05-03 \\
\end{flushright}

\vspace{1.5cm}

\begin{center}
  {\Large\bf Automated computation of one-loop integrals in massless theories\\
  }
  \vspace{1cm}
  {\large Andr\'e van Hameren, Jens Vollinga and Stefan Weinzierl\\
  \vspace{1cm}
      {\small \em Institut f{\"u}r Physik, Universit{\"a}t Mainz,}\\
      {\small \em D - 55099 Mainz, Germany}\\
  } 
\end{center}

\vspace{2cm}

\begin{abstract}\noindent
  {
   We consider one-loop tensor and scalar integrals, 
   which occur in a massless quantum field theory and
   we report on the implementation into a numerical program
   of an algorithm for the automated
   computation of these one-loop integrals. 
   The number of external legs of the loop integrals is not restricted.
   All calculations are done within dimensional regularization. 
   }
\end{abstract}

\vspace*{\fill}

\newpage

\section{Introduction}
\label{sect:intro}

Jet physics plays an important r\^ole at the TEVATRON and will become
even more important at the LHC. 
It provides information on the strong interactions and forms quite often
important backgrounds for searches of new physics.
While jet observables can rather easily be modelled at leading order (LO)
in perturbation 
theory \cite{Berends:1987me,Berends:1989ie,Berends:1990ax,Caravaglios:1995cd,Caravaglios:1998yr,Draggiotis:1998gr,Draggiotis:2002hm,Stelzer:1994ta,Pukhov:1999gg,Yuasa:1999rg}, 
this description suffers from several drawbacks.
A leading order calculation depends strongly on the renormalization scale
and can therefore give only an order-of-magnitude-estimate on absolute rates.
Secondly, at leading order a jet is modelled by a single parton. This is a very crude
approximation and oversimplifies inter- and intra-jet correlations.
The situation is improved by including higher-order corrections in perturbation theory.

At present, there are many NLO calculation for $2 \rightarrow 2$ processes at
hadron colliders, but only a few for $2 \rightarrow 3$ processes.
Fully differential numerical programs exist for example for
$p p \rightarrow \mbox{3 jets}$ \cite{Kilgore:1996sq,Nagy:2001fj,Nagy:2003tz},
$p p \rightarrow V + \mbox{2 jets}$ \cite{Campbell:2002tg},
$p p \rightarrow t \bar{t} H$ \cite{Beenakker:2002nc,Dawson:2003zu} and
$p p \rightarrow H + \mbox{2 jets}$ \cite{DelDuca:2001eu,DelDuca:2001fn}.
The NLO calculation for $p p \rightarrow t \bar{t} + \mbox{jet}$ is in progress \cite{Brandenburg:2004fw}.
In the examples cited above the relevant one-loop amplitudes were
usually calculated in an hand-crafted way by a mixture of analytical and numerical
methods.
However it has become clear, that this traditional way reaches its limits when the number of 
external particles increases.
On the other hand, it is desirable to have NLO calculations for $2 \rightarrow n$ processes in hadron-hadron
collisions with $n$ in the range of $n=3,4,...,6,7$.
QCD processes like $p p \rightarrow n \; \mbox{jets}$ form often important backgrounds for the searches of
signals of new physics.
To overcome the computational limitation, there were in the last years
several proposals for the automated computation 
of one-loop amplitudes
\cite{Soper:1998ye,Soper:1999xk,Passarino:2001wv,Ferroglia:2002mz,Nagy:2003qn,Denner:2002ii,Dittmaier:2003bc,Giele:2004iy,Giele:2004ub,delAguila:2004nf,Pittau:2004bc,Binoth:2002xh,Binoth:2004tn,vanHameren:2004wr}.

In this paper we report on the implementation of an algorithm for the automated
computation of one-loop integrals, which occur in a massless quantum field theory
into a numerical program.
For QCD processes at high-energy colliders the massless approximation is justified 
for all quarks except the top quark.
The number of external particles of the loop integrals is not restricted within our approach.
All calculations are done within dimensional regularization. 
When combined with the appropriate contributions coming from the emission of an additional parton, the project
we report on here will provide a numerical program for the automated computation of $2 \rightarrow n$ NLO processes
in massless QCD. 
As our approach is valid (in theory) for all $n$, the actual 
limitation on $n$ will result from the available computer power for the Monte Carlo integration.

The problem which we address in this paper is the fast and efficient numerical evaluation
of scalar and tensor one-loop integrals in a massless quantum field theory.
Tensor integrals are loop integrals, where the loop momentum also appears in the
numerator.
Loop integrals are classified according to the number $n$ of internal propagators
(or equivalently the number of external legs), as well as the rank $r$, counting
the power to which the loop momentum occurs in the numerator.
It is a well known fact, that all one-loop integrals can be expressed in terms
of the scalar two-, three- and four-point functions, up to some trivial extra integrals,
which are mainly related to a specific choice of the regularization scheme.
The task is to calculate numerically the coefficients in front of the basic scalar integrals.
It is tempting to do this with a single algorithm, which covers all cases in a uniform way.
Although several of these algorithms exist, a particular algorithm will perform well for most
configurations, but can lead to numerical instabilities in certain corners of configuration space.
We therefore opted for a ``patch-work''-style, treating loop integrals with
$n$ propagators and rank $r$ on an individual
basis. This reduces to a certain extent the dependency on the caveats of a particular algorithm
and allows us rather easily to replace in future releases of the program a particular reduction method
with an improved version.

We employed the following strategies for the reduction of one-loop integrals:
The two-point functions are rather easy and are therefore evaluated directly.
For the reduction of tensor integrals with $n\ge3$ we use spinor methods
and follow mainly the recent work by del Aguila and Pittau 
\cite{Pittau:1997ez,Pittau:1998mv,Weinzierl:1998we,delAguila:2004nf}.
This leads to scalar integrals, where additional powers of the $\eps$-components
of the loop momentum can still be present in the numerator.
If such powers are present, the resulting integrals are rather easy and are evaluated directly.
It remains to treat scalar $n$-point integrals with $n \ge5$ and to reduce them to the basic set.
For $n=5$ and $n=6$ the reduction is unique
\cite{Bern:1994kr,Binoth:1999sp,Fleischer:1999hq,Denner:2002ii}.
This is no longer true for $n\ge7$. In the latter case we use a method based on the singular
value decomposition of the Gram matrix
\cite{Duplancic:2003tv,Giele:2004iy}.
These steps reduce all integrals to the basic set of scalar two-, three- and four-point functions.
The latter are then evaluated in terms of logarithms and dilogarithms.

This paper is organized as follows:
In the next section we introduce our notation.
Section \ref{sect:tensor} discusses the reduction of tensor integrals.
Section \ref{sect:highdim} evaluates higher-dimensional integrals, resulting from
additional powers of the $\eps$-components
of the loop momentum in the numerator.
Section \ref{sect:scalar} treats the reduction of scalar $n$-point integrals for $n \ge 5$.
In section \ref{sect:impl} we comment on the numerical implementation.
Finally, section \ref{sect:concl} contains our conclusions.
In an appendix we provide the necessary details on spinors as well as the explicit
expressions for the basic scalar integrals and methods for the numerical evaluation
of some special functions.


\section{Definitions and conventions}
\label{sect:def}

\begin{figure}
\begin{center}
\begin{picture}(300,200)(0,0)
\Vertex(110,31){2}
\Vertex(110,169){2}
\Vertex(190,31){2}
\Vertex(190,169){2}
\Vertex(70,100){2}
\Vertex(230,100){2}
\ArrowLine(110,31)(70,100)
\ArrowLine(70,100)(110,169)
\ArrowLine(110,169)(190,169)
\ArrowLine(190,169)(230,100)
\ArrowLine(230,100)(190,31)
\ArrowLine(190,31)(110,31)
\ArrowLine(110,31)(101,13)
\ArrowLine(70,100)(50,100)
\ArrowLine(110,169)(101,187)
\ArrowLine(190,169)(199,187)
\ArrowLine(230,100)(250,100)
\ArrowLine(190,31)(199,13)
\Text(99,11)[tr]{$p_1$}
\Text(46,100)[r]{$p_2$}
\Text(150,189)[b]{...}
\Text(254,100)[l]{$p_{n-1}$}
\Text(201,11)[tl]{$p_{n}$}
\Text(150,37)[b]{$k_n$}
\Text(94,66)[lb]{$k_1$}
\Text(94,134)[lt]{$k_2$}
\Text(206,66)[rb]{$k_{n-1}$}
\end{picture}
\caption{\label{figure1}
The labelling for a generic one-loop integral. The arrows denote the momentum flow.}
\end{center}
\end{figure}
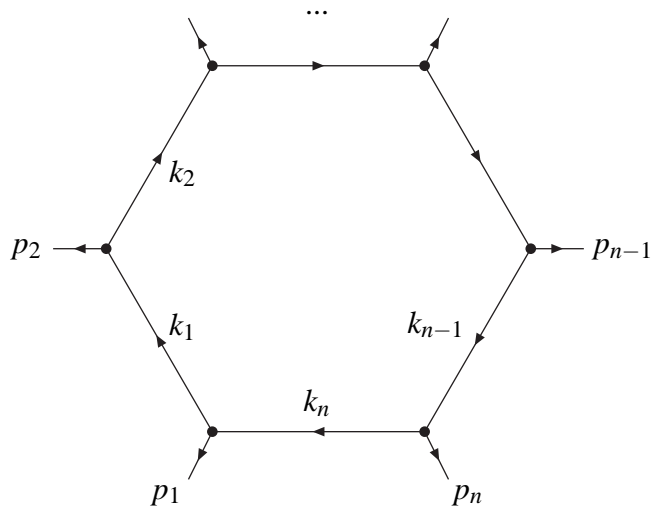

The general convention for a scalar one-loop $n$-point integral is
\bq
I_n & = &
 e^{\eps \gamma_E} \mu^{2\eps} \int \frac{d^Dk}{i \pi^{\frac{D}{2}}}
  \frac{1}{k^2 (k-p_1)^2 ... (k-p_1-...p_{n-1})^2},
\eq
with $D=4-2\eps$.
We further use the notation
\bq
 q_i = \sum\limits_{j=1}^i p_j, & & k_i = k - q_i. 
\eq
The flow of momentum is shown in fig. (\ref{figure1}).
The kinematical matrix $S$ is defined by
\bq
 S_{ij} & = & \left( q_i - q_j \right)^2,
\eq
and the Gram matrix is defined by
\bq
G_{ij} & = & 2 q_i q_j.
\eq
Integrals of the type
\bq
I_n^{\mu_1 ... \mu_r} & = &
 e^{\eps \gamma_E} \mu^{2\eps} \int \frac{d^Dk}{i \pi^{\frac{D}{2}}}
  \frac{k^{\mu_1} ... k^{\mu_r}}{k^2 (k-p_1)^2 ... (k-p_1-...p_{n-1})^2}
\eq
are called tensor integrals. These integrals are said to have rank $r$, if the
loop momentum appears $r$-times in the numerator.
These integrals are always contracted with a coefficient $J^n_{\mu_1 ... \mu_r}$, 
which is a product of $n$ tree-level currents.
This coefficient depends on the momenta and the polarization vectors of the external particles
of the scattering process.
Since trees can be attached to the external lines of a one-loop integral, the external
momenta $p_j$ of a one-loop integral are in general not the momenta of the external
particles in the scattering process, but rather sums of the latter.
The coefficient $J^n_{\mu_1 ... \mu_r}$ can be computed efficiently in four dimensions.

It is therefore appropriate to discuss different variants of
dimensional regularization.
The most commonly used schemes are the conventional dimensional
regularization scheme (CDR) \cite{Collins}, 
where all momenta and all polarization vectors are taken to be in $D$ dimensions,
the 't Hooft-Veltman scheme (HV) \cite{'tHooft:1972fi}, 
where the momenta and the helicities of the unobserved particles are $D$ dimensional,
whereas the momenta and the helicities of the observed particles are 4 dimensional,
and the four-dimensional helicity scheme (FD) \cite{Bern:1992aq,Weinzierl:1999xb,Bern:2002zk}, 
where all polarization vectors are kept in four dimensions, as well
as the momenta of the observed particles. 
Only the momenta of the unobserved particles are continued to $D$ dimensions.

The conventional scheme is mostly used for an analytical calculation of the interference of a one-loop
amplitude with the Born amplitude by using polarization sums corresponding to $D$ dimensions.
For the calculation of one-loop helicity amplitudes the 't Hooft-Veltman scheme
and the four-dimensional helicity scheme are possible choices.
All schemes have in common, that the propagators appearing in the denominator of the
loop-integrals are continued to $D$ dimensions. They differ how they treat the algebraic
part in the numerator.
In the 't Hooft-Veltman scheme the algebraic part is treated in $D$ dimensions, whereas
in the FD scheme the algebraic part is treated in four dimensions.
It is possible to relate results obtained in one scheme to another scheme, using simple
and universal transition formulae \cite{Kunszt:1994sd,Signer:PhD,Catani:1997pk}.

Since the efficient numerical calculation of the coefficient $J^n_{\mu_1 ... \mu_r}$
relies on the Fierz identity in four dimensions, we are lead to the choice of the
four-dimensional helicity scheme.
In this scheme we can assume without loss of generality that the 
coefficient $J^n_{\mu_1 ... \mu_r}$
is given by
\bq
 J^n_{\mu_1 ... \mu_r} & = &
  \left\l a_1 - \left| \gamma_{\mu_1} \right| b_1 - \right\r
  ...
  \left\l a_r - \left| \gamma_{\mu_r} \right| b_r - \right\r,
\eq
where $\l a_i - |$ and $| b_j - \r$ are Weyl spinors of definite helicity.
It is convenient to denote spinor inner products as follows:
\bq
\left\l p q \right\r = \left\l p - | q + \right\r,
 & &
\left[ q p \right] = \left\l q + | p - \right\r.
\eq
Important relations satisfied by the Weyl spinors are
\bq
\left\l p - \left| \gamma_\mu \right| q - \right\r & = & 
 \left\l q + \left| \gamma_\mu \right| p +\right\r,
\eq
and the Fierz identity
\bq
\left\l a - \left| \gamma_\mu \right| b - \right\r \left\l c + \left| \gamma^\mu \right| d +\right\r
 & = & 2 \left\l a d \right\r \left[ c b \right].
\eq
Therefore we consider tensor integrals of the form
\bq
\label{basictensorintegral1}
I_n^r & = &
 e^{\eps \gamma_E} \mu^{2\eps} 
  \left\l a_1 - \left| \gamma_{\mu_1} \right| b_1 - \right\r
  ...
  \left\l a_r - \left| \gamma_{\mu_r} \right| b_r - \right\r
  \int \frac{d^Dk}{i \pi^{\frac{D}{2}}}
  \frac{k^{\mu_1}_{(4)} ... k^{\mu_r}_{(4)}}{k^2 (k-p_1)^2 ... (k-p_1-...p_{n-1})^2},
 \nonumber \\
\eq
where $k^{\mu}_{(4)}$ denotes the projection of the $D$ dimensional vector $k^\mu$ onto
the four-dimensional subspace.
A peculiarity of the four-dimensional helicity scheme is given by the fact that the dot
product of $k^{\mu}_{(4)}$ with itself does not cancel exactly a propagator, i.e.
the $D$-dimensional $k_{(D)}^2$ is given as the sum of the four-dimensional $k_{(4)}^2$
and $k_{(-2\eps)}^2$, consisting of the $\eps$-components:
\bq
k_{(D)}^2 & = & k_{(4)}^2 + k_{(-2\eps)}^2.
\eq
When no conflicting interpretations are possible, we will often drop the indication of the dimension
of the underlying space.
As a consequence we have to consider a generalization of eq. (\ref{basictensorintegral1}) 
by allowing additional powers of $k_{(-2\eps)}^2$ in the numerator:
\bq
\label{basictensorintegral2}
I_n^{r,s} & = &
 e^{\eps \gamma_E} \mu^{2\eps} 
  \left\l a_1 - \left| \gamma_{\mu_1} \right| b_1 - \right\r
  ...
  \left\l a_r - \left| \gamma_{\mu_r} \right| b_r - \right\r
  \int \frac{d^Dk}{i \pi^{\frac{D}{2}}}
  \frac{\left(-k_{(-2\eps)}^2\right)^s \; k^{\mu_1}_{(4)} ... k^{\mu_r}_{(4)}}{k^2 (k-p_1)^2 ... (k-p_1-...p_{n-1})^2},
 \nonumber \\
\eq
The result of eq. (\ref{basictensorintegral2}) can be expressed in the form
\bq
I_n^{r,s} & = &
 \frac{C_{-2}}{\eps^2} + \frac{C_{-1}}{\eps} + C_0
 + {\cal O}(\eps).
\eq
We are mainly interested in the coefficient $C_0$. Besides that, the knowledge of the
coefficients $C_{-2}$ and $C_{-1}$ provides additional cross checks, as the divergent part
of the Laurent series has to cancel against similar parts coming from the real emission and
renormalization.
The purpose of this paper is to set up a scheme for the numerical calculation of the 
coefficients $C_{-2}$, $C_{-1}$ and $C_0$.


\section{Tensor reduction}
\label{sect:tensor}

The classical method for the reduction of tensor one-loop integrals is the Passarino-Veltman algorithm
\cite{Passarino:1979jh,Stuart:1988tt,Stuart:1990de,Devaraj:1998es}.
Here, we use instead spinor methods, discussed for example in 
\cite{Pittau:1997ez,Pittau:1998mv,Weinzierl:1998we,delAguila:2004nf}.
The spinor methods have the advantage that they avoid to a large extent the appearance
of Gram determinants, or in cases where they cannot be avoided, reduce them to square roots of
Gram determinants.
Alternative methods, like for example approaches based on dual vectors or raising and lowering
operators are discussed in 
\cite{vanOldenborgh:1990wn,Davydychev:1991va,Tarasov:1996br,Tarasov:1997kx,Campbell:1997zw}.
In this section we give an algorithm for the reduction of integrals 
of the form as in eq. (\ref{basictensorintegral2})
towards integrals of the form
\bq
\label{basicscalar1}
I_n^{0 ,s} & = &
 e^{\eps \gamma_E} \mu^{2\eps} 
  \int \frac{d^Dk}{i \pi^{\frac{D}{2}}}
  \frac{\left(-k_{(-2\eps)}^2\right)^s }{k^2 (k-p_1)^2 ... (k-p_1-...p_{n-1})^2}.
\eq
We do this by treating the different cases of $n$ separately: Two-point functions ($n=2$) are rather simple
and are calculated directly in section \ref{subsect:tensortwopoint}.
For $n \ge 3$ we use spinor methods. The cases $n=3$ and $n=4$ are special,
as there are only two, respectively three independent external momenta.
The tensor three-point functions are discussed in section \ref{subsect:tensortwopoint},
while the tensor four-point functions are treated in section \ref{subsect:tensorfourpoint}.
Finally, for $n \ge 5$ we use a general method for the tensor reduction, which is discussed
in section \ref{subsect:tensorfivepoint}.

\subsection{Generalities}

The basic formula for the Passarino-Veltman algorithm states, that the scalar product
of a loop momentum with an external momentum reduces the rank of the tensor integral:
\bq
\label{basicpassarinoveltman}
2 p_i \cdot k_j & = & k_{i-1}^2 - k_i^2 + q_i^2 - q_{i-1}^2 - 2 p_i q_j.
\eq
This formula is valid independently of which variant of dimensional regularization
is used, since the $\eps$-components cancel between $k_{i-1}^2$  and $k_i^2$.
Therefore the subscript indicating if the loop momentum $k_j$ lives in $D$ or four
dimensions was dropped.

The first step for the construction of the reduction algorithm based on spinor methods
is to associate to each $n$-point loop integral a pair of two light-like momenta $l_1$ and
$l_2$, which are linear combinations of two external momenta $p_i$ and $p_j$ of the loop
integral under consideration \cite{delAguila:2004nf}.
Note that $p_i$ and $p_j$ need not be light-like.
Obviously, this construction only makes sense for three-point integrals and beyond, 
as for two-point integrals there is only one independent external momentum.
If $p_i$ and $p_j$ are light-like, the construction of $l_1$ and $l_2$ is trivial:
\bq
l_1 = p_i, 
& &
l_2 = p_j.
\eq
If $p_i$ is light-like, but $p_j$ is massive one has
\bq
l_1 = p_i, 
& &
l_2 = -\alpha_2 p_i + p_j,
\eq
where
\bq
\alpha_2 & = & \frac{p_j^2}{2p_i p_j}.
\eq
The inverse formula is given by
\bq
p_i = l_1, 
& &
p_j = \alpha_2 l_1 + l_2.
\eq
If both $p_i$ and $p_j$ are massive,
one has
\bq
l_1 = \frac{1}{1-\alpha_1 \alpha_2} \left( p_i - \alpha_1 p_j \right), 
& &
l_2 = \frac{1}{1-\alpha_1 \alpha_2} \left( -\alpha_2 p_i + p_j \right).
\eq
If $2p_ip_j>0$, $\alpha_1$ and $\alpha_2$ are given by
\bq
\alpha_1 = \frac{2p_ip_j-\sqrt{\Delta}}{2p_j^2},
 & &
\alpha_2 = \frac{2p_ip_j-\sqrt{\Delta}}{2p_i^2}.
\eq
If $2p_ip_j<0$, we have the formulae
\bq
\alpha_1 = \frac{2p_ip_j+\sqrt{\Delta}}{2p_j^2},
 & &
\alpha_2 = \frac{2p_ip_j+\sqrt{\Delta}}{2p_i^2}.
\eq
Here,
\bq
\Delta & = & \left( 2p_ip_j \right)^2 - 4p_i^2 p_j^2.
\eq
The signs are chosen in such away that the light-like limit $p_i^2 \rightarrow 0$ (or $p_j^2 \rightarrow 0$)
is approached smoothly.
The inverse formula is given by
\bq
\label{decompmom}
p_i = l_1 + \alpha_1 l_2, 
& &
p_j = \alpha_2 l_1 + l_2.
\eq
Note that $l_1$, $l_2$ are real for $\Delta>0$.
For $\Delta<0$, $l_1$ and $l_2$ acquire imaginary parts.
These formulae can be used in the following ways:
First we may decompose any four-vector $p$ into a sum of two null-vectors:
\bq
p & = & \alpha n + l,
\eq
where $n$ is an arbitrary null-vector and
\bq 
l = -\alpha n + p, & & \alpha = \frac{p^2}{2pn}.
\eq
Secondly, we may decompose $k\!\!\!/_l$ as follows:
\bq
\label{basiceqn}
k\!\!\!/_l & = & 
 \frac{1}{2l_1l_2} 
   \left[
         \left( 2k_l l_2 \right) l\!\!\!/_1 + \left( 2 k_l l_1 \right) l\!\!\!/_2
         - l\!\!\!/_1 k\!\!\!/_l l\!\!\!/_2
         - l\!\!\!/_2 k\!\!\!/_l l\!\!\!/_1
   \right],
\eq
where $l_1$ and $l_2$ are obtained from decomposing $p_i$ and $p_j$ into null-vectors.
Note that this formula can be proved by solely using the anti-commutation relations for the Dirac matrices
and is therefore valid in the HV/CDR-scheme as well as in the FD-scheme.
The main application for eq. (\ref{basiceqn}) will be the application towards
the spinor strings
\bq
\label{basiceqn2}
  \left\l a - \left| k\!\!\!/_{(4)} \right| b - \right\r
 & =  &
        \frac{1}{2l_1l_2} 
   \left[
         \left( 2k l_2 \right) \left\l a - \left| l\!\!\!/_1 \right| b - \right\r
         + \left( 2 k l_1 \right) \left\l a - \left| l\!\!\!/_2 \right| b - \right\r
 \right.
 \nonumber \\
 & & 
 \left.
         - \left\l a l_1 \right\r \left[ l_2 b \right] \left\l l_2 - \left| k\!\!\!/_{(4)} \right| l_1 - \right\r
         - \left\l a l_2 \right\r \left[ l_1 b \right] \left\l l_1 - \left| k\!\!\!/_{(4)} \right| l_2 - \right\r
   \right],
\eq
appearing in eq. (\ref{basictensorintegral2}).
We note that the scalar products of $k_l$ with $l_1$ or $l_2$ are linear combinations of the scalar products
of $k_l$ with $p_i$ and $p_j$, 
\bq
\label{l1orl2scalprod}
2 k_l l_1 = \frac{1}{1-\alpha_1\alpha_2} \left[ 2p_i k_l - \alpha_1 2p_j k_l \right],
 & &
2 k_l l_2 = \frac{1}{1-\alpha_1\alpha_2} \left[ -\alpha_2 2p_i k_l + 2p_j k_l \right],
\eq
and therefore immediately reduce the rank of the tensor integral through eq. (\ref{basicpassarinoveltman}).
Eq. (\ref{basiceqn2}) allows us to replace an arbitrary sandwich
\bq
  \left\l a - \left| k\!\!\!/_{(4)} \right| b - \right\r
\eq
with the standard types 
\bq
 \left\l l_1 - \left| k\!\!\!/_{(4)} \right| l_2 - \right\r
 & \mbox{and} & 
 \left\l l_2 - \left| k\!\!\!/_{(4)} \right| l_1 - \right\r,
\eq
plus additional reduced integrals. This procedure can easily be iterated.
Note that $l_1$ and $l_2$ depend on the external momenta of the loop integral.
In general, these two vectors have to be re-defined when pinching a propagator.

\subsection{The tensor two-point function}
\label{subsect:tensortwopoint}

The tensor two-point function is special, as it does not fit into the general scheme, which we use for the
tensor reduction. This is due to the fact that the two-point function depends only on one external momentum.
Fortunately, the two-point function is simple enough, such that one can solve the problem by direct calculation.
We consider the general tensor two-point integral 
\bq
I_2^{\mu_1 ... \mu_r,s} & = &
 e^{\eps \gamma_E} \mu^{2\eps} \int \frac{d^Dk}{i \pi^{\frac{D}{2}}}
  \left(-k_{(-2\eps)}^2\right)^s \frac{k^{\mu_1} ... k^{\mu_r}}{(k-p)^2 k^2 }
 \\
 & = &
 e^{\eps \gamma_E} \mu^{2\eps} \int\limits_0^1 da \int \frac{d^Dk}{i \pi^{\frac{D}{2}}}
   \left(-k_{(-2\eps)}^2\right)^s
   \left( k+a p \right)^{\mu_1} ... \left( k+a p \right)^{\mu_r} 
   \left[ -k^2 + a(1-a)\left(-p^2\right) \right]^{-2}.
 \nonumber
\eq
Expanding $\left( k+a p \right)^{\mu_1} ... \left( k+a p \right)^{\mu_r}$ yields terms
of the form
\bq
a^{r-2t} k^{\mu_{\sigma(1)}} ... k^{\mu_{\sigma(2t)}} p^{\mu_{\sigma(2t+1)}} ... p^{\mu_{\sigma(r)}}.
\eq
Note that terms with an odd number of $k^{\mu}$'s vanish after integration.
We further have
\bq
\int \frac{d^Dk}{i \pi^{\frac{D}{2}}} k^{\mu_1} k^{\mu_2} f(k^2)
 & = & 
  \frac{g^{\mu_1 \mu_2}}{D} 
  \int \frac{d^Dk}{i \pi^{\frac{D}{2}}} k^2 f(k^2),
 \nonumber \\
\int \frac{d^Dk}{i \pi^{\frac{D}{2}}} k^{\mu_1} k^{\mu_2} k^{\mu_3} k^{\mu_4} f(k^2)
 & = & 
  \frac{ g^{\mu_1 \mu_2} g^{\mu_3 \mu_4}
        +g^{\mu_1 \mu_3} g^{\mu_2 \mu_4}
        +g^{\mu_1 \mu_4} g^{\mu_2 \mu_3} 
       }{D(D+2)} 
  \int \frac{d^Dk}{i \pi^{\frac{D}{2}}} \left( k^2 \right)^2 f(k^2).
 \nonumber 
\eq
In general we have
\bq
\int \frac{d^Dk}{i \pi^{\frac{D}{2}}} k^{\mu_1} ... k^{\mu_{2w}} f(k^2)
 & = & 
  2^{-w} \frac{\Gamma\left(\frac{D}{2}\right)}{\Gamma\left(\frac{D}{2}+w\right)}
  \left( g^{\mu_1 \mu_2} ... g^{\mu_{2w-1} \mu_{2w}} + \mbox{permutations}
                         \right)
  \int \frac{d^Dk}{i \pi^{\frac{D}{2}}} \left( k^2 \right)^w f(k^2).
 \nonumber
\eq
The fully symmetric tensor structure 
\bq
 S^{\mu_1 .... \mu_{2w}} & = & g^{\mu_1 \mu_2} ... g^{\mu_{2w-1} \mu_{2w}} + \mbox{permutations}
\eq
has $(2w-1)!!=(2w-1)(2w-3)...1$ terms.
We obtain in the absence of powers of $k_{(-2\eps)}^2$
\bq
\label{scal_two_point_no_eps}
\lefteqn{
 e^{\eps \gamma_E} \mu^{2\eps} \int\limits_0^1 da \; a^{r-2t}
   \int \frac{d^Dk}{i \pi^{\frac{D}{2}}}
   k^{\mu_1} ... k^{\mu_{2t}}
   \left[ -k^2 + a(1-a)\left(-p^2\right) \right]^{-2}
 = } & &\nonumber \\
& = &
 \left(-\frac{p^2}{2} \right)^t S^{\mu_1 .... \mu_{2t}}
 \frac{\Gamma(1+r-t-\eps) \Gamma(2-2\eps)}
      {\Gamma(1-\eps) \Gamma(2+r-2\eps)} I_2
 \nonumber \\
& = &
 \left( - \frac{p^2}{2} \right)^t S^{\mu_1 .... \mu_{2t}}
 \frac{(r-t)!}{(r+1)!} \left\{ 1 + \eps \left[ 2 Z_1(r+1) - Z_1(r-t) - 2 \right] + {\cal O}(\eps^2) \right\} I_2,
\eq
where $Z_1(n)$ is a harmonic sum
\bq
 Z_1(n) & = & \sum\limits_{j=1}^n \frac{1}{j},
\eq
and $I_2$ is the scalar two-point function:
\bq
I_2 & = & 
 e^{\eps \gamma_E} \left( \frac{-p^2}{\mu^2} \right)^{-2\eps}
  \frac{\Gamma(-\eps) \Gamma(1-\eps)^2}{\Gamma(2-2\eps)}
 = 
 \frac{1}{\eps} + 2 - \ln\left( \frac{-p^2}{\mu^2} \right) + {\cal O}(\eps).
\eq
Since $I_2$ starts at $1/\eps$ we can neglect ${\cal O}(\eps^2)$ terms in eq. (\ref{scal_two_point_no_eps}).
If powers of $k_{(-2\eps)}^2$ are present, we obtain if all indices are contracted into four-dimensional quantities
\bq
\lefteqn{
 e^{\eps \gamma_E} \mu^{2\eps} \int\limits_0^1 da \; a^{r-2t}
   \int \frac{d^Dk}{i \pi^{\frac{D}{2}}}
   \left(-k_{(-2\eps)}^2\right)^s
   k^{\mu_1} ... k^{\mu_{2t}}
   \left[ -k^2 + a(1-a)\left(-p^2\right) \right]^{-2}
 = } & &\nonumber \\
& = &
 - \eps \left( p^2 \right)^s \left( - \frac{p^2}{2} \right)^t 
  S^{\mu_1 .... \mu_{2t}}
 \frac{(s-1)! (r+s-t)!}{(r+2s+1)!} I_2 + {\cal O}(\eps)
 \nonumber \\
 & &
 + \mbox{terms, which vanish when contracted into $4$-dimensional quantities}.
\eq

\subsection{The tensor three-point function}
\label{subsect:tensorthreepoint}

For tensor three-point integrals we may use eq. (\ref{basiceqn2}).
The first two terms on the r.h.s. of eq. (\ref{basiceqn2}) reduce the rank immediately.
We can therefore assume that the tensor structure is a product of
\bq
\l l_1- | k\!\!\!/_l^{(4)} | l_2- \r \;\;\;\mbox{and}\;\;\; \l l_2- | k\!\!\!/_l^{(4)} | l_1- \r.
\eq
Note that the index of the loop momentum is irrelevant,
\bq
\l l_1- | k\!\!\!/_1^{(4)} | l_2- \r = \l l_1- | k\!\!\!/_2^{(4)} | l_2- \r 
                                     = \l l_1- | k\!\!\!/_3^{(4)} | l_2- \r,
\eq
since the following sandwiches vanish:
\bq
\l l_1- | p\!\!\!/_1 | l_2- \r = \l l_1- | p\!\!\!/_2 | l_2- \r = 0.
\eq
Two different spinor types reduce the rank:
\bq
\label{differentspinortype}
\l l_1- | k\!\!\!/_l^{(4)} | l_2- \r \l l_2- | k\!\!\!/_l^{(4)} | l_1- \r
 & = & 
  \left( 2l_1 k_l \right) \left( 2 l_2 k_l \right) 
  - \left( 2 l_1 l_2 \right) \left(k_l^{(4)}\right)^2.
\eq
Here, $\left(k_l^{(4)}\right)^2$ 
denotes the square of the four-dimensional components and does
not exactly cancel a propagator.
Using
\bq
 k_{(4)}^2 & = & k_{(D)}^2 - k_{(-2\eps)}^2
\eq
together with eq. (\ref{higherdimint}) will lead to integrals in higher dimensions.
It remains to discuss the case of
a tensor structure of the same spinor type, e.g. either
\bq
\label{trisametype}
\l l_1- | k\!\!\!/_l^{(4)} | l_2- \r  ... \l l_1- | k\!\!\!/_l^{(4)} | l_2- \r,
\eq
or the same situation with $l_1$ and $l_2$ exchanged.
It is easy to see that these terms will vanish after integration, since any contraction of
\bq
\l l_1- | \gamma_{\mu_1} | l_2- \r  ... \l l_1- | \gamma_{\mu_r} | l_2- \r
\eq
with $p_1^\mu$, $p_2^\mu$ or $g^{\mu \nu}$ will vanish.

\subsection{The tensor four-point function}
\label{subsect:tensorfourpoint}

For the four-point function two new features appear: 
One can no longer shift freely the loop momentum inside the spinor sandwiches
and tensor structures of the same spinor type, as in eq. (\ref{trisametype}), no longer vanish identically.
On the other hand, the four-point function has,
apart from the two external momenta $p_i$ and $p_j$ used to construct $l_1$ and $l_2$,
one additional independent external momentum, labelled $p_3$ in the following.
For the tensor reduction one starts again with eq. (\ref{basiceqn2}), possibly preceded by a shift
in the loop momentum, which synchronizes all occuring loop momenta in the numerator from
\bq
 k^{\mu_1}_{l_1} ... k^{\mu_r}_{l_r}
 & \mbox{to} &
 k^{\mu_1}_{l} ... k^{\mu_r}_{l}.
\eq
It is therefore sufficient to consider a tensor structure, which is a product of
\bq
\label{startfourpoint}
\l l_1- | k\!\!\!/_l^{(4)} | l_2- \r \;\;\;\mbox{and}\;\;\; \l l_2- | k\!\!\!/_l^{(4)} | l_1- \r.
\eq
If in the tensor structure both spinor types appear, we can use eq. (\ref{differentspinortype}): 
\bq
\l l_1- | k\!\!\!/_l^{(4)} | l_2- \r \l l_2- | k\!\!\!/_l^{(4)} | l_1- \r
 & = & 
  \left( 2l_1 k_l \right) \left( 2 l_2 k_l \right) 
  - \left( 2 l_1 l_2 \right) \left(k_l^{(4)}\right)^2.
 \nonumber
\eq
If on the other hand in the tensor structure only a single spinor type occurs, we now
use the third external momentum $p_3$ and write:
\bq
\label{endfourpoint}
\lefteqn{
\l l_1- | k\!\!\!/_l^{(4)} | l_2- \r \l l_1- | k\!\!\!/_l^{(4)} | l_2- \r
 = 
 - 
 \frac{\l l_1- | p\!\!\!/_3 | l_2- \r}{\l l_2- | p\!\!\!/_3 | l_1- \r}
      \l l_1- | k\!\!\!/_l^{(4)} | l_2- \r \l l_2- | k\!\!\!/_l^{(4)} | l_1- \r
} & & \nonumber \\
 & & 
 + \frac{\l l_1- | k\!\!\!/_l^{(4)} | l_2- \r}{\l l_2- | p\!\!\!/_3 | l_1- \r}
 \left[ 
        \left( 2 l_1 p_3 \right) \left( 2 l_2 k_l\right) 
      + \left( 2 l_2 p_3 \right) \left( 2 l_1 k_l\right) 
      - \left( 2 l_1 l_2 \right) \left( 2 p_3 k_l\right) 
 \right].
\eq
For the first term one uses in turn again eq. (\ref{differentspinortype}), while the last term
in the square bracket reduces the rank by one through eq. (\ref{basicpassarinoveltman})
and eq. (\ref{l1orl2scalprod}).
This allows to reduce any rank $r \ge 2$ integral to scalar or rank $1$ integrals.
It remains to treat rank $1$ integrals. For rank 1 integrals we may use
\bq
\l l_1- | k\!\!\!/_l^{(4)} | l_2- \r  & = & 
  \frac{1}{\l l_2- | p\!\!\!/_3 | l_1- \r} \mbox{Tr}_+ \left( k\!\!\!/_l^{(4)} l\!\!\!/_2 p\!\!\!/_3 l\!\!\!/_1 \right),
 \nonumber \\
\l l_2- | k\!\!\!/_l^{(4)} | l_1- \r  & = & 
  \frac{1}{\l l_1- | p\!\!\!/_3 | l_2- \r} \mbox{Tr}_+ \left( k\!\!\!/_l^{(4)} l\!\!\!/_1 p\!\!\!/_3 l\!\!\!/_2 \right),
\eq
where the subscript ``$+$'' indicates that a projection operator $(1+\gamma_5)/2$ has been inserted into
the trace.
Since the piece proportional to the totally antisymmetric tensor vanishes after
integration, we may replace
\bq
 \mbox{Tr}_+ \left( k\!\!\!/_l^{(4)} l\!\!\!/_2 p\!\!\!/_3 l\!\!\!/_1 \right) 
 \rightarrow 
 \frac{1}{2} \mbox{Tr}\left( k\!\!\!/_l^{(4)} l\!\!\!/_2 p\!\!\!/_3 l\!\!\!/_1 \right),
& &
 \mbox{Tr}_+ \left( k\!\!\!/_l^{(4)} l\!\!\!/_1 p\!\!\!/_3 l\!\!\!/_2 \right) 
 \rightarrow 
 \frac{1}{2} \mbox{Tr}\left( k\!\!\!/_l^{(4)} l\!\!\!/_1 p\!\!\!/_3 l\!\!\!/_2 \right).
 \nonumber \\
\eq
Therefore
\bq
\l l_1- | k\!\!\!/_l^{(4)} | l_2- \r  & = & 
  \frac{1}{2 \l l_2- | p\!\!\!/_3 | l_1- \r} 
  \left[ ( 2 l_1 p_3 ) ( 2 l_2 k_l ) + ( 2 l_2 p_3 ) ( 2 l_1 k_l )
                          - ( 2 l_1 l_2 ) ( 2 p_3 k_l ) \right]
 \nonumber \\
 & & + \; \mbox{terms, which vanish after integration},
 \nonumber \\
\l l_2- | k\!\!\!/_l^{(4)} | l_1- \r  & = & 
  \frac{1}{2 \l l_1- | p\!\!\!/_3 | l_2- \r} 
  \left[ ( 2 l_1 p_3 ) ( 2 l_2 k_l ) + ( 2 l_2 p_3 ) ( 2 l_1 k_l )
                          - ( 2 l_1 l_2 ) ( 2 p_3 k_l ) \right]
 \nonumber \\
 & & + \; \mbox{terms, which vanish after integration}.
\eq
This allows to reduce rank $1$ four-point integrals.

\subsection{The tensor five-point function and beyond}
\label{subsect:tensorfivepoint}

Here we discuss the tensor reduction of $n$-point functions with $n \ge 5$.
For rank $r \ge 2$ we follow the same steps in eq. (\ref{startfourpoint}) - eq. (\ref{endfourpoint}) 
as for the four-point function.
The only difference occurs in the treatment of rank one integrals.
We note that for the $n$-point functions with $n \ge 5$ we have 
one further additional independent momentum, which will be labelled $p_4$.
For the rank one integrals we have \cite{delAguila:2004nf}
\bq
\lefteqn{
\l l_1- | k\!\!\!/_l^{(4)} | l_2- \r  = 
         - \frac{1}{\delta} \left[  \left( 2 l_1 p_4 \right) \left( 2 l_2 k_l \right)
                                  + \left( 2 l_2 p_4 \right) \left( 2 l_1 k_l \right)
                                  - \left( 2 l_1 l_2 \right) \left( 2 p_4 k_l \right)
                            \right] \l l_1- | p\!\!\!/_3 | l_2- \r
 } \nonumber \\
 & &
         + \frac{1}{\delta} \left[  \left( 2 l_1 p_3 \right) \left( 2 l_2 k_l \right)
                                  + \left( 2 l_2 p_3 \right) \left( 2 l_1 k_l \right)
                                  - \left( 2 l_1 l_2 \right) \left( 2 p_3 k_l \right)
                            \right] \l l_1- | p\!\!\!/_4 | l_2- \r,
 \hspace*{20mm}
 \nonumber \\
\lefteqn{
\l l_2- | k\!\!\!/_l^{(4)} | l_1- \r  =  
         \frac{1}{\delta} \left[  \left( 2 l_1 p_4 \right) \left( 2 l_2 k_l \right)
                                  + \left( 2 l_2 p_4 \right) \left( 2 l_1 k_l \right)
                                  - \left( 2 l_1 l_2 \right) \left( 2 p_4 k_l \right)
                            \right] \l l_2- | p\!\!\!/_3 | l_1- \r
 } \nonumber \\
 & &
         - \frac{1}{\delta} \left[  \left( 2 l_1 p_3 \right) \left( 2 l_2 k_l \right)
                                  + \left( 2 l_2 p_3 \right) \left( 2 l_1 k_l \right)
                                  - \left( 2 l_1 l_2 \right) \left( 2 p_3 k_l \right)
                            \right] \l l_2- | p\!\!\!/_4 | l_1- \r,
\eq
where
\bq
\delta & = & 
 \left\l l_1 - \left| p\!\!\!/_4 \right| l_2 - \right\r
 \left\l l_2 - \left| p\!\!\!/_3 \right| l_1 - \right\r
 -
 \left\l l_1 - \left| p\!\!\!/_3 \right| l_2 - \right\r
 \left\l l_2 - \left| p\!\!\!/_4 \right| l_1 - \right\r.
\eq
$\delta$ is proportional to the square root of the Gram determinant of the four-momenta $l_1$, $l_2$, $p_3$ and
$p_4$. Numerical instabilities in the limit $\delta \rightarrow 0$ can be treated with the methods discussed in 
ref. \cite{delAguila:2004nf}.


\section{Higher-dimensional integrals}
\label{sect:highdim}

In this section we discuss the evaluation of scalar integrals of the form
\bq
I_n^{0 ,s} & = &
 e^{\eps \gamma_E} \mu^{2\eps} 
  \int \frac{d^Dk}{i \pi^{\frac{D}{2}}}
  \frac{\left(-k_{(-2\eps)}^2\right)^s }{k^2 (k-p_1)^2 ... (k-p_1-...p_{n-1})^2},
\eq
with $s>0$. Scalar integrals with $s=0$ are treated in section \ref{sect:scalar}.
In a space of $D=2m-2\eps$ dimensions (with $m$ being an integer), we decompose $k_{(D)}^2$ as follows:
\bq
k_{(D)}^2 & = & k_{(2m)}^2 + k_{(-2\eps)}^2
\eq
If a power of $(-k_{(-2\eps)}^2)$ appears in the numerator we have
\cite{Bern:1995db}
\bq
\label{higherdimint}
\int \frac{d^{2m-2\varepsilon}k}{\pi^{m-\varepsilon}i}
  (-k_{(-2\eps)}^2)^s f(k_{(2m)}^\mu,k_{(-2\eps)}^2) 
 & = &
   \frac{\Gamma(s-\eps)}{\Gamma(-\eps)}
   \int \frac{d^{2m+2s-2\varepsilon}k}{\pi^{m+s-\varepsilon}i} f(k_{(2m)}^\mu,k_{(-2\eps)}^2).
\eq
The effect of a factor of $(-k_{(-2\eps)}^2)^s$ in the numerator is to shift the dimension by $2s$.
Note that $\Gamma(s-\eps)/\Gamma(-\eps)$ brings an explicit factor of $\eps$, therefore we have to
take higher-dimen\-sional integrals into account only if they are divergent.
A scalar $n$-point integral with unit powers of the propagators is finite, 
if \cite{Giele:2004iy}
\bq
2 < \frac{D}{2} < n.
\eq
Here $2<D/2$ is the condition to be infrared finite and $D/2<n$ is the condition to be UV-finite.
Therefore, higher dimensional integrals are always infrared finite and we only have to
calculate the UV-pole of the higher dimensional integrals.
This can easily be done. For $m \ge n$ we find
\bq
I_n & = &
 e^{\eps \gamma_E} \mu^{2\eps} \int \frac{d^{2m-2\eps}k}{i \pi^{m-\eps}}
  \frac{1}{k_1^2 k_2^2 ... k_n^2}
 \nonumber \\
 & = &
 \frac{1}{\eps} 
 \frac{(-1)^m}{(m-n)!} \int d^na \; \delta\left(1-\sum\limits_{j=1}^n a_j \right)
  {\cal F}^{m-n}
 + {\cal O}\left( \eps^0 \right),
\eq
where
\bq
 {\cal F} & = &
   - \sum\limits_{i<j} a_i a_j \left(p_{i+1} + ... + p_j \right)^2.
\eq
Note that the integral over the Feynman parameters is a polynomial in the Feynman parameters and can
be done according to the formula
\bq
 \int d^na \; \delta\left(1-\sum\limits_{j=1}^n a_j \right)
  a_1^{\nu_1-1} ... a_n^{\nu_n-1}
 & = & 
 \frac{\Gamma(\nu_1) ... \Gamma(\nu_n)}{\Gamma(\nu_1+...+\nu_n)}.
\eq
In practice there are additional simplifications:
When calculating one-loop amplitudes, we are free to choose an appropriate gauge.
Using the Feynman gauge, we can ensure that the rank $r$ of a loop integral
is always less or equal the number of external legs $n$. In addition, there
are obviously no powers of $k_{(-2\eps)}^2$ in the original loop integral, i.e. we have
\bq
r \le n
 \;\;\; \mbox{and} \;\;\;
 s = 0.
\eq
The algorithm for the tensor reduction in section \ref{sect:tensor}
respects the inequality
\bq
r + 2 s \le n. 
\eq
Therefore the only non-zero higher-dimensional integrals which occur in the Feynman gauge result from
the two-point function with a single power of $k_{(-2\eps)}^2$ in the numerator ($n=2$ and $s=1$),
the three-point function with a single power of $k_{(-2\eps)}^2$ in the numerator ($n=3$ and $s=1$)
and the four-point function with two powers of $k_{(-2\eps)}^2$ in the numerator ($n=4$ and $s=2$).
The case of the two-point function has already been 
discussed explicitly in section \ref{subsect:tensortwopoint}.
For the remaining two cases one finds:
\bq
 e^{\eps \gamma_E} \mu^{2\eps} \int \frac{d^{D}k}{i \pi^{\frac{D}{2}}}
  \frac{\left(-k_{(-2\eps)}^2\right)}{k_1^2 k_2^2 k_3^2}
 & = &
 \frac{1}{2} + {\cal O}(\eps),
 \nonumber \\
 e^{\eps \gamma_E} \mu^{2\eps} \int \frac{d^{D}k}{i \pi^{\frac{D}{2}}}
  \frac{\left(-k_{(-2\eps)}^2\right)^2}{k_1^2 k_2^2 k_3^2 k_4^2}
 & = &
 - \frac{1}{6} + {\cal O}(\eps).
\eq


\section{Reduction of higher point scalar integrals}
\label{sect:scalar}

In this section we discuss the reduction of scalar integrals of the form
\bq
I_n & = &
 e^{\eps \gamma_E} \mu^{2\eps} 
  \int \frac{d^Dk}{i \pi^{\frac{D}{2}}}
  \frac{1}{k^2 (k-p_1)^2 ... (k-p_1-...p_{n-1})^2},
\eq
with $n \ge 5$ to a basic set of scalar two-, three- and four-point functions.
It is a long known fact, that higher point scalar integrals can be expressed
in terms of this basic set \cite{Melrose:1965kb,vanNeerven:1984vr}, however
the practical implementation within dimensional regularization
was only worked out 
recently \cite{Bern:1994kr,Binoth:1999sp,Fleischer:1999hq,Denner:2002ii,Duplancic:2003tv}.
We distinguish three different cases: Scalar pentagons (i.e. scalar five-point functions),
scalar hexagons (scalar six-point functions) and scalar integrals with more than
six propagators.

\subsection{Reduction of pentagons}

A five-point function in $D=4-2\eps$ dimensions can be expressed as a sum of four-point functions, where
one propagator is removed, plus a five-point function in $6-2\eps$ dimensions \cite{Bern:1994kr}.
Since the $(6-2\eps)$-dimensional pentagon is finite and comes with an extra factor of $\eps$ in front, it does
not contribute at ${\cal O}(\eps^0)$. In detail we have
\bq
I_5 & = & -2\eps B I_5^{6-2\eps}
          + \sum\limits_{i=1}^5 b_i I_4^{(i)}
 =
          \sum\limits_{i=1}^5 b_i I_4^{(i)}
  + {\cal O}\left(\eps\right),
\eq
where $I_5^{6-2\eps}$ denotes the $(6-2\eps)$-dimensional pentagon and
$I_4^{(i)}$ denotes the four-point function, which is obtained from the pentagon by removing propagator $i$.
The coefficients $B$ and $b_i$ are obtained from the kinematical matrix $S_{ij}$ as follows: 
\bq
b_i = \sum\limits_j \left( S^{-1} \right)_{ij},
 & &
B = \sum\limits_{i} b_i.
\eq

\subsection{Reduction of hexagons}

The six-point function can be expressed as a sum of five-point functions \cite{Binoth:1999sp}
\bq
\label{scalarsixpoint}
I_6 & = & \sum\limits_{i=1}^6 b_i I_5^{(i)}.
\eq
The coefficients $b_i$ are again related to the kinematical matrix $S_{ij}$: 
\bq
b_i & = & \sum\limits_j \left( S^{-1} \right)_{ij}.
\eq

\subsection{Reduction of scalar integrals with more than six propagators}

For the seven-point function and beyond we can again express the $n$-point function as a sum over
$(n-1)$-point functions \cite{Duplancic:2003tv}: 
\bq
\label{scalarnpoint}
I_n & = & \sum\limits_{i=1}^n r_i I_{n-1}^{(i)}.
\eq
In contrast to eq. (\ref{scalarsixpoint}), the decomposition in eq. (\ref{scalarnpoint}) is no longer unique.
A possible set of coefficients $r_i$ can be
obtained from the singular value decomposition of the $(n-1) \times (n-1)$ Gram matrix 
\bq
G_{ij} & = & \sum\limits_{k=1}^4 U_{ik} w_k \left(V^T\right)_{kj}.
\eq
as follows \cite{Giele:2004iy}
\bq
  r_i & = & \frac{V_{i 5}}{W_5}, \;\;\; 1 \le i \le n-1,
 \nonumber \\
  r_n & = & - \sum\limits_{j=1}^{n-1} r_j,
\eq
with
\bq
W_5 & = & \frac{1}{2} \sum\limits_{j=1}^{n-1} G_{j j} V_{j 5}.
\eq
Note that the kernel of $G_{ij}$ is spanned by the vectors
$V_{i 5}$, $V_{i 6}$, ..., $V_{i (n-1)}$.


\section{Numerical implementation}
\label{sect:impl}

We have implemented the algorithms described so far into a numerical computer program.
The program is able to calculate the coefficients $C_{-2}$, $C_{-1}$ and $C_0$
of the Laurent expansion of one-loop $n$-point integrals of rank $r$ and $s$ powers of 
$k_{(-2\eps)}^2$ in the numerator:
\bq
I_n^{r,s} & = &
 \frac{C_{-2}}{\eps^2} + \frac{C_{-1}}{\eps} + C_0
 + {\cal O}(\eps).
\eq
As our algorithms are valid for any number of $n$ external particles, the actual limitation
on $n$ will result from the available computer power.

We have performed several checks on our computer code.
The value of a tensor integral 
\bq
I_n^r & = &
 e^{\eps \gamma_E} \mu^{2\eps} 
  \left\l a_1 - \left| \gamma_{\mu_1} \right| b_1 - \right\r
  ...
  \left\l a_r - \left| \gamma_{\mu_r} \right| b_r - \right\r
  \int \frac{d^Dk}{i \pi^{\frac{D}{2}}}
  \frac{k^{\mu_1}_{(4)} ... k^{\mu_r}_{(4)}}{k^2 (k-p_1)^2 ... (k-p_1-...p_{n-1})^2}
 \nonumber 
\eq
is clearly unchanged
if we permute the tensor structure
\bq
  \left\l a_1 - \left| \gamma_{\mu_1} \right| b_1 - \right\r
  ...
  \left\l a_r - \left| \gamma_{\mu_r} \right| b_r - \right\r
 \rightarrow
  \left\l a_{\sigma(1)} - \left| \gamma_{\mu_1} \right| b_{\sigma(1)} - \right\r
  ...
  \left\l a_{\sigma(r)} - \left| \gamma_{\mu_r} \right| b_{\sigma(r)} - \right\r.
\eq
Since our algorithm reduces the rank step by step, this actually provides a non-trivial check.

Secondly, for specific choices of the tensor structure, like
\bq
 \left\l p_i - \left| k\!\!\!/_j \right| p_i - \right\r & = & 2 p_i k_j,
\eq
the numerator reduces immediately to simpler integrals. This will lead to relations among
different integrals, which can be checked numerically.

Finally, we have written three independent codes (in two different programming languages: Fortran and C++),
which all agree with each other.

For future reference we give a few numerical results.
We start by specifying a set of twelve light-like momenta $p_i$, with $i=1,...,12$.
This serves as the input data for the scalar 12-point function, where all external particles 
are light-like.
By combining four-vectors we can obtain the external kinematics of lower point functions.
We choose the set
\bq
 \left\{ p_1, ..., p_{j-1}, p_j+...+p_{12} \right\}
\eq
for the $j$-point function. For $3 \le j < 12$ this corresponds to $j-1$ light-like external legs and one
massive leg. Note that for $j=2$ we have a two-point function with light-like external momenta, which vanishes.
The random values for our set of momenta (in units of GeV) are:
{\small
\bq
  p_1 & = & (5.897009121257959,-1.971772490149703,-4.63646682189329,-3.064311543033953),
  \nonumber \\
  p_2 & = & (9.78288114803946,-3.495678805323657,-7.42828599660035,-5.320297021726135),
  \nonumber \\
  p_3 & = & (3.751716626791747,0.3633444560526895,-2.74701214525531,2.529285023049251),
  \nonumber \\
  p_4 & = & (14.8572007649265,-9.282840702083684,9.182091233148681,7.08903968497886),
  \nonumber \\
  p_5 & = & (4.056006277332882,-1.236594041315223,0.8781947326421281,3.761754392608195),
  \nonumber \\
  p_6 & = & (2.023022829577847,0.3217130479853592,0.6516721562887716,1.887973909901054),
  \nonumber \\
  p_7 & = & (23.51469894530697,20.57030957903025,3.67304549050126,-10.78481187299925),
  \nonumber \\
  p_8 & = & (6.161822860155142,0.9716060205020823,2.082735149413637,5.717189606638176),
  \nonumber \\
  p_9 & = & (10.67981737238498,-2.237405231711613,-2.487945529176884,-10.14212226215274),
  \nonumber \\
  p_{10} & = & (9.275824054226526,-4.002681832986503,0.83197173093136,8.326300082736525),
  \nonumber \\
  p_{11} & = & (-45,0,0,45),
  \nonumber \\
  p_{12} & = & (-45,0,0,-45).
\eq
}
This set satisfies momentum conservation
\bq
 \sum\limits_{j=1}^{12} p_j & = & 0.
\eq
We give values for tensor integrals up to rank $2$. Since for higher rank integrals no new reduction
algorithms are used, this is sufficient for demonstration purposes.
For $j$-point integrals with $j\le 10$ the momenta $p_{11}$ and $p_{12}$ have no special relation to the
external kinematics (only the sum $p_j+...+p_{12}$ corresponds to an external leg).
Therefore the sandwich
\bq
 \left\l p_{12} - \left| k\!\!\!/_1 \right| p_{11} - \right\r
\eq
is an example of a generic rank $1$ integral.
Similar, we use for $j\le 8$ the tensor structure 
\bq
 \left\l p_{12} - \left| k\!\!\!/_1 \right| p_{11} - \right\r
 \left\l p_{10} - \left| k\!\!\!/_1 \right| p_{9} - \right\r.
\eq
The numerical values of the bra- and ket-spinors depends on a choice for the phases of the spinors.
Our conventions are listed
in the appendix. In addition we use, when evaluating spinors, a rotation 
$(x,y,z) \rightarrow (z,x,y)$ 
for the spatial coordinates of a four-vector, such that the line, where spinors are not defined, lies along
the negative $y$-axis. This avoids problems with incoming particles, which are often taken to be on the $z$-axis.
For cross-checks we also quote the numerical values of the spinors in our convention:
{\small
\bq
 \left\l p_{12} - \right| & = & (-6.708203932499369, -6.708203932499369),
 \nonumber \\
 \left\l p_{10} - \right| & = & (3.179275984427568, 2.618929631626706+1.258991623436299 i ),
 \nonumber \\
 \left| p_{11} - \right\r & = & (6.708203932499369, -6.708203932499369),
 \nonumber \\
 \left| p_{9} - \right\r & = & (2.862144623041976, -3.543539407653475-0.7817233321122782 i ).
\eq
}
The results of the loop integrals will depend also on the renormalization scale $\mu$. We set
\bq
 \mu & = & 135 \; \mbox{GeV}.
\eq
This specifies all input parameters.
The results for the coefficients $C_{-2}$, $C_{-1}$ and $C_0$ of the Laurent expansion 
are shown for the scalar integrals 
in table \ref{resrank0}.
The corresponding numbers for the rank $1$ integrals can be found in table \ref{resrank1},
while table \ref{resrank2} shows the results for the rank 2 integrals.
Our independent programs agree within $10^{-7}$.
Table \ref{cputime} shows the CPU time 
in seconds for a tensor integral with $n$ external legs and
rank $r$ for $r \le n \le 10$ on a standard PC equipped with a Pentium IV running at 2 GHz.
The recursive algorithm is efficiently implemented with the help of look-up tables.
The required memory for the look-up tables is negligible, i.e. of the order of 
$10 \;\mbox{MB}$ for the case $n=r=10$.
\begin{table}
\begin{center}
\begin{tabular}{|c|rrr|}
\hline
 $n$ & $C_{-2}$ & $C_{-1}$ & $C_0$ \\
\hline
  3 & $ 9.4327 \cdot 10^{0}    $ & $ ( 1.1371 + 0.2963 i ) \cdot 10^{2}   $ & $ ( 6.3106 + 3.5723 i ) \cdot 10^{2}   $ \\
  4 & $ 3.0405 \cdot 10^{-1}   $ & $ ( 3.8038 + 0.9552 i ) \cdot 10^{0}   $ & $ ( 2.2164 + 1.1950 i ) \cdot 10^{1}   $ \\
  5 & $ 1.3863 \cdot 10^{-3}   $ & $ ( 1.8815 + 0.4355 i ) \cdot 10^{-2}  $ & $ ( 1.2211 + 0.5911 i ) \cdot 10^{-1}  $ \\
  6 & $ 6.7736 \cdot 10^{-6}   $ & $ ( 9.8029 + 2.1280 i ) \cdot 10^{-5}  $ & $ ( 6.8667 + 3.0797 i ) \cdot 10^{-4}  $ \\
  7 & $ 3.1190 \cdot 10^{-8}   $ & $ ( 4.6404 + 0.9799 i ) \cdot 10^{-7}  $ & $ ( 3.3624 + 1.4578 i ) \cdot 10^{-6}  $ \\
  8 & $ 1.68789 \cdot 10^{-11} $ & $ ( 2.4470 + 0.5303 i ) \cdot 10^{-10} $ & $ ( 1.7380 + 0.7688 i ) \cdot 10^{-9}  $ \\
  9 & $ 9.56014 \cdot 10^{-15} $ & $ ( 1.2782 + 0.3003 i ) \cdot 10^{-13} $ & $ ( 8.4151 + 4.0157 i ) \cdot 10^{-13} $ \\
 10 & $ 4.41232 \cdot 10^{-18} $ & $ ( 5.3676 + 1.3862 i ) \cdot 10^{-17} $ & $ ( 3.1673 + 1.6863 i ) \cdot 10^{-16} $ \\
 11 & $ 1.51680 \cdot 10^{-21} $ & $ ( 1.7511 + 0.4765 i ) \cdot 10^{-20} $ & $ ( 9.6086 + 5.5011 i ) \cdot 10^{-20} $ \\
 12 & $-8.17311 \cdot 10^{-25} $ & $ (-9.3478 - 2.5677 i ) \cdot 10^{-24} $ & $ (-5.0527 - 2.9367 i ) \cdot 10^{-23} $ \\
\hline
\end{tabular}
\caption{\label{resrank0}
Results for the scalar $n$-point functions with $3 \le n \le 12$. The $C_i$ denote the coefficients of the Laurent
series.
}
\end{center}
\end{table}
\begin{table}
\begin{center}
\begin{tabular}{|c|rrr|}
\hline
 $n$ & $C_{-2}$ & $C_{-1}$ & $C_0$ \\
\hline
  3 & $ 0                                    $ & $ ( 2.3701 + 1.2937 i ) \cdot 10^{3}   $ & $ ( 2.9247 + 2.5629 i ) \cdot 10^{4} $ \\
  4 & $ (-1.0164 - 0.4783 i ) \cdot 10^{2}   $ & $ (-1.1710 - 0.9390 i ) \cdot 10^{3}   $ & $ (-6.1191 - 7.9185 i ) \cdot 10^{3} $ \\
  5 & $ (-7.6639 - 3.6047 i ) \cdot 10^{-1}  $ & $ (-9.7028 - 7.5033 i ) \cdot 10^{0}   $ & $ (-5.7693 - 6.8698 i ) \cdot 10^{1} $ \\
  6 & $ (-4.3386 - 2.0356 i ) \cdot 10^{-3}  $ & $ (-5.7717 - 4.3739 i ) \cdot 10^{-2}  $ & $ (-3.6478 - 4.1735 i ) \cdot 10^{-1} $ \\
  7 & $ (-2.0606 - 0.9984 i ) \cdot 10^{-5}  $ & $ (-2.7898 - 2.1356 i ) \cdot 10^{-4}  $ & $ (-1.8100 - 2.0589 i ) \cdot 10^{-3} $ \\
  8 & $ (-1.0155 - 0.6865 i ) \cdot 10^{-8}  $ & $ (-1.3143 - 1.2571 i ) \cdot 10^{-7}  $ & $ (-0.8305 - 1.1134 i ) \cdot 10^{-6} $\\
  9 & $ (-4.0783 - 6.1625 i ) \cdot 10^{-12} $ & $ (-4.1976 - 8.6804 i ) \cdot 10^{-11} $ & $ (-2.1879 - 6.2292 i ) \cdot 10^{-10} $\\
 10 & $ (-1.0266 - 3.9800 i ) \cdot 10^{-15} $ & $ (-0.2878 - 4.7984 i ) \cdot 10^{-14} $ & $ ( 0.2754 - 2.8456 i ) \cdot 10^{-13} $ \\
\hline
\end{tabular}
\caption{\label{resrank1}
Results for the $n$-point functions of rank 1. The $C_i$ denote the coefficients of the Laurent
series.
}
\end{center}
\end{table}
\begin{table}
\begin{center}
\begin{tabular}{|c|rrr|}
\hline
 $n$ & $C_{-2}$ & $C_{-1}$ & $C_0$ \\
\hline
 3 & $ 0                                   $ & $ ( 3.1317 + 4.3445 i ) \cdot 10^{5}  $ & $ ( 3.3251 + 7.4740 i ) \cdot 10^{6} $\\
 4 & $ (-0.7042 - 1.0292 i ) \cdot 10^{4}  $ & $ (-0.6161 - 1.5979 i ) \cdot 10^{5}  $ & $ (-0.1596 - 1.1662 i ) \cdot 10^{6} $ \\
 5 & $ (-5.3188 - 7.7592 i ) \cdot 10^{1}  $ & $ (-0.5233 - 1.2881 i ) \cdot 10^{3}  $ & $ (-0.1818 - 1.0225 i ) \cdot 10^{4} $\\
 6 & $ (-3.0368 - 4.3882 i ) \cdot 10^{-1} $ & $ (-3.1500 - 7.5262 i ) \cdot 10^{0}  $ & $ (-1.2159 - 6.2003 i ) \cdot 10^{1} $ \\
 7 & $ (-1.4611 - 2.0679 i ) \cdot 10^{-3} $ & $ (-1.5537 - 3.6078 i ) \cdot 10^{-2} $ & $ (-0.6331 - 3.0290 i ) \cdot 10^{-1} $\\
 8 & $ (-6.5351 - 9.7049 i ) \cdot 10^{-7} $ & $ (-0.7106 - 1.7046 i ) \cdot 10^{-5} $ & $ (-0.2959 - 1.4465 i ) \cdot 10^{-4} $\\
\hline
\end{tabular}
\caption{\label{resrank2}
Results for the $n$-point functions of rank 2. The $C_i$ denote the coefficients of the Laurent
series.
}
\end{center}
\end{table}
\begin{table}
\begin{center}
\begin{tabular}{|cc|lllllllllll|}
\hline
 & r & 0 & 1 & 2 & 3 & 4 & 5 & 6 & 7 & 8 & 9 & 10 \\
 n & & & & & & & & & & & & \\
\hline
 2 & & $1 \cdot 10^{-6}$ & $5 \cdot 10^{-6}$ & $2 \cdot 10^{-5}$ & & & & & & & & \\
 3 & & $1 \cdot 10^{-6}$ & $2 \cdot 10^{-5}$ & $2 \cdot 10^{-4}$ & $1 \cdot 10^{-3}$ & & & & & & & \\
 4 & & $2 \cdot 10^{-6}$ & $5 \cdot 10^{-5}$ & $4 \cdot 10^{-4}$ & $2 \cdot 10^{-3}$ & $6 \cdot 10^{-3}$ & & & & & & \\
 5 & & $3 \cdot 10^{-5}$ & $1 \cdot 10^{-4}$ & $6 \cdot 10^{-4}$ & $3 \cdot 10^{-3}$ & $9 \cdot 10^{-3}$ & $0.03$ & & & & & \\
 6 & & $2 \cdot 10^{-4}$ & $3 \cdot 10^{-4}$ & $9 \cdot 10^{-4}$ & $4 \cdot 10^{-3}$ & $0.02$ & $0.04$ & $0.1$ & & & & \\
 7 & & $7 \cdot 10^{-4}$ & $7 \cdot 10^{-4}$ & $1 \cdot 10^{-3}$ & $5 \cdot 10^{-3}$ & $0.02$ & $0.06$ & $0.1$ & $0.4$ & & & \\
 8 & & $3 \cdot 10^{-3}$ & $3 \cdot 10^{-3}$ & $4 \cdot 10^{-3}$ & $8 \cdot 10^{-3}$ & $0.02$ & $0.07$ & $0.2$ & $0.6$ & $1.8$ & & \\
 9 & & $0.01$ & $0.01$ & $0.01$ & $0.02$ & $0.03$ & $0.08$ & $0.3$ & $0.9$ & $2.6$ & $7$ & \\
10 & & $0.05$ & $0.05$ & $0.05$ & $0.06$ & $0.06$ & $0.2$ & $0.4$ & $1.1$ & $3.5$ & $8$ & $25$ \\
\hline
\end{tabular}
\caption{\label{cputime}
CPU time in seconds for a tensor integral with $n$ external legs and
rank $r$ for $r \le n$ on a standard PC ( Pentium IV with 2 GHz).
}
\end{center}
\end{table}


\section{Conclusions}
\label{sect:concl}

In this paper we discussed an algorithm for the automated computation of one-loop integrals, which occur
in a massless quantum field theory.
This is relevant for high-energy experiments, where the masses of the quarks (with the exception of the top quark)
can usually be neglected.
We reported on the implementation of this algorithm into a numerical program.
It is worth to point out, that there are a priori no restrictions on the number of external legs of the loop integrals.
Therefore the actual restriction is only given by the available computer resources.
We gave examples for the evaluation of loop integrals with up to twelve external legs.
In future work we intend to integrate this program into a package for the automatic calculation of jet cross sections.


\begin{appendix}

\section{Complex four-vectors and spinors}

In this appendix we list our conventions for spinors.
In particular we comment on complex four-vectors and their associated spinors.
Although the external momenta of a loop integral are real quantities, the decomposition
of two massive vectors into linear combinations of null-vectors, as in 
eq. (\ref{decompmom}), may introduce complex four-vectors.
For the metric we use
\bq
g_{\mu \nu} & = & \mbox{diag}(+1,-1,-1,-1).
\eq
A null-vector satisfies
\bq
\left( p_0 \right)^2 - \left( p_1 \right)^2 - \left( p_2 \right)^2 - \left( p_3 \right)^2 & = & 0.
\eq
This relation holds also for complex $p_\mu$.
Light-cone coordinates are as follows:
\bq
p_+ = p_0 + p_3, \;\;\; p_- = p_0 - p_3, \;\;\; p_{\bot} = p_1 + i p_2,
                                         \;\;\; p_{\bot^\ast} = p_1 - i p_2.
\eq
Note that $p_{\bot^\ast}$ does not involve a complex conjugation of $p_1$ or $p_2$.
We use the Weyl representation for the Dirac matrices
\bq
\gamma^{\mu} = \left(\begin{array}{cc}
 0 & \sigma^{\mu} \\
 \bar{\sigma}^{\mu} & 0 \\
\end{array} \right),
& &
\gamma_{5} = i \gamma^0 \gamma^1 \gamma^2 \gamma^3 
           = \left(\begin{array}{cc}
 1 & 0 \\
 0 & -1 \\
\end{array} \right),
\eq
where the 4-dimensional $\sigma^{\mu}$-matrices are
\bq
\sigma_{A \dot{B}}^{\mu} = \left( 1 , - \vec{\sigma} \right), & &
\bar{\sigma}^{\mu \dot{A} B} = \left( 1 ,  \vec{\sigma} \right),
\eq
and Pauli matrices $\vec{\sigma} = (\sigma_x, \sigma_y, \sigma_z)$ are as usual
\bq
\sigma_x = \left(\begin{array}{cc}
 0 & 1\\
 1 & 0 \\
\end{array} \right),
&
\sigma_y = \left(\begin{array}{cc}
 0 & -i\\
 i & 0 \\
\end{array} \right),
&
\sigma_z = \left(\begin{array}{cc}
 1 & 0\\
 0 & -1 \\
\end{array} \right).
\eq
Four-component Dirac spinors are constructed out of two Weyl spinors:
\bq
\label{Weylket}
u(p) & = & \left(\begin{array}{c} \left| p + \right\r \\ \left| p - \right\r \\ \end{array} \right) 
       =   \left(\begin{array}{c} p_A \\ p^{\dot{B}} \\ \end{array} \right)
       =   \left(\begin{array}{c} u_+(p) \\ u_-(p) \\ \end{array} \right).
\eq
where
\bq
u_\pm(p) & = & \frac{1}{2} \left( 1 \pm \gamma_5 \right)u(p).
\eq
Bra-spinors are given by
\bq
\label{Weylbra}
\overline{u}(p) & = & \left( \left\l p - \right|, \left\l p + \right| \right)
                  =   \left( p^A, p_{\dot{B}} \right)
                  =   \left( \bar{u}_-(p), \bar{u}_+(p) \right),
\eq
where
\bq
\bar{u}_\pm(p) & = & \bar{u}(p) \frac{1}{2} \left( 1 \mp \gamma_5 \right).
\eq
Eq. (\ref{Weylket}) and eq. (\ref{Weylbra}) show three different notations for Weyl spinors.
We are using mainly the bra-ket notation.
In terms of the light-cone components of a null-vector, the corresponding spinors can be chosen
as
\bq
\left| p+ \right\r = \frac{1}{\sqrt{\left| p_+ \right|}} \left( \begin{array}{c}
  -p_{\bot^\ast} \\ p_+ \end{array} \right),
 & &
\left| p- \right\r = \frac{e^{-i \phi}}{\sqrt{\left| p_+ \right|}} \left( \begin{array}{c}
  p_+ \\ p_\bot \end{array} \right),
 \nonumber \\
\left\l p+ \right| = \frac{e^{-i\phi}}{\sqrt{\left| p_+ \right|}} 
 \left( -p_\bot, p_+ \right),
 & &
\left\l p- \right| = \frac{1}{\sqrt{\left| p_+ \right|}} 
 \left( p_+, p_{\bot^\ast} \right),
\eq
where the phase $\phi$ is given by
\bq
p_+ & = & \left| p_+ \right| e^{i\phi}.
\eq
The spinor products are then given by
\bq
\left\l p q \right\r & = & \left\l p- | q+ \right\r
  = \frac{1}{\sqrt{\left| p_+ \right| \left| q_+ \right|}} 
    \left( p_{\bot^\ast} q_+ - p_+ q_{\bot^\ast} \right),
 \nonumber \\
\left[ q p \right] & = & \left\l q+ | p- \right\r
  = \frac{1}{\sqrt{\left| p_+ \right| \left| q_+ \right|}} e^{-i\phi_p} e^{-i\phi_q}
    \left( p_\bot q_+ - p_+ q_\bot \right).
\eq


\section{The basic scalar integrals}

In this appendix we list the basic scalar integrals, which are the scalar two-point, the scalar three-point and the
scalar four-point functions in $D=4-2\eps$ dimensions.
Since we restrict ourselves to massless quantum field theories, all internal propagators are massless and we only
have to distinguish the masses of the external momenta.
All scalar integrals have been known for a long time in the literature.
Classical papers on scalar integrals are \cite{'tHooft:1979xw,Denner:1991qq}.
Scalar integrals within dimensional regularization are treated in \cite{Bern:1993em,Bern:1994kr}.
Useful information on the three-mass triangle can be found in \cite{Ussyukina:1993jd,Lu:1992ny,Bern:1997ka}.
The scalar boxes have been recalculated in \cite{Duplancic:2000sk,Duplancic:2002dh}.

\subsection{The two-point function}

The scalar two-point function is given by
\bq
I_2(p_1^2,\mu^2) & = & 
 \frac{1}{\eps} + 2 - \ln\left(\frac{-p_1^2}{\mu^2}\right) 
 + {\cal O}(\eps).
\eq

\subsection{Three-point functions}

For the three-point functions we have three different cases: One external mass, two external masses
and three external masses.
The one-mass scalar triangle with $p_1^2\neq 0$, $p_2^2=p_3^2=0$ is given by
\bq
I^{1m}_3(p_1^2,\mu^2) & = &
 \frac{1}{\eps^2 p_1^2} 
 - \frac{1}{\eps p_1^2} \ln\left(\frac{-p_1^2}{\mu^2}\right)
 + \frac{1}{2 p_1^2} \ln^2\left(\frac{-p_1^2}{\mu^2}\right)
 - \frac{1}{2 p_1^2} \zeta_2
 + {\cal O}(\eps).
\eq
The two-mass scalar triangle with $p_1^2\neq 0$, $p_2^2\neq 0$ and $p_3^2=0$ is given by
\bq
I^{2m}_3(p_1^2,p_2^2,\mu^2) & = & 
 \frac{1}{\eps} 
 \frac{1}{\left( p_1^2-p_2^2 \right)}
 \left[ - \ln\left(\frac{-p_1^2}{\mu^2}\right) + \ln\left(\frac{-p_2^2}{\mu^2}\right) \right]
\nonumber \\
 & &
 + \frac{1}{2(p_1^2-p_2^2)} 
   \left[ \ln^2\left(\frac{-p_1^2}{\mu^2}\right) - \ln^2\left(\frac{-p_2^2}{\mu^2}\right) \right]
 + {\cal O}(\eps).
\eq
The three-mass scalar triangle with $p_1^2\neq 0$, $p_2^2\neq 0$ and $p_3^2\neq 0$:
This integral is finite and we have
\bq
 I_3^{3m}\left(p_1^2,p_2^2,p_3^2,\mu^2\right) & = &
    - \int\limits_0^1 d^3 \alpha 
      \frac{\delta\left(1-\alpha_1-\alpha_2-\alpha_3\right)}
           {-\alpha_1\alpha_2 p_1^2 - \alpha_2 \alpha_3 p_2^2 - \alpha_3 \alpha_1 p_3^2}
 + {\cal O}(\eps).
\eq
With the notation
\bq
 & &
 \delta_1 = p_1^2 - p_2^2 - p_3^2,
 \;\;\;
 \delta_2 = p_2^2 - p_3^2 - p_1^2,
 \;\;\;
 \delta_3 = p_3^2 - p_1^2 - p_2^2,
 \nonumber \\
 & &
 \Delta_3 = \left(p_1^2\right)^2 + \left(p_2^2\right)^2 + \left(p_3^2\right)^2 
               - 2 p_1^2 p_2^2 - 2 p_2^2 p_3^2 - 2 p_3^2 p_1^2,
\eq
the three-mass triangle $I_3^{3m}$ is expressed in the region
$p_1^2,p_2^2,p_3^2 < 0$ and $\Delta_3 < 0$  by 
\bq
\lefteqn{
 I_3^{3m} = 
   -\frac{2}{\sqrt{-\Delta_3}} 
} & & \nonumber \\
 & & 
   \times \left[
   \mbox{Cl}_2\left( 2 \arctan \left( \frac{\sqrt{-\Delta_3}}{\delta_1}\right)\right)
   +\mbox{Cl}_2\left( 2 \arctan \left( \frac{\sqrt{-\Delta_3}}{\delta_2}\right)\right) 
   +\mbox{Cl}_2\left( 2 \arctan \left( \frac{\sqrt{-\Delta_3}}{\delta_3}\right)\right)
   \right]
 \nonumber \\
 & & + {\cal O}(\eps).
\eq
The Clausen function $\mbox{Cl}_2(x)$ is defined in eq. (\ref{defclausen}).
In  the region $p_1^2,p_2^2,p_3^2 < 0$ and $\Delta_3 > 0$ as well as in the region
$p_1^2, p_3^2 < 0$, $p_2^2 > 0$ (for which $\Delta_3$ is always positive) the integral
$I_3^{3m}$ is given by 
\bq
 I_3^{3m} & = & \frac{1}{\sqrt{\Delta_3}} \mbox{Re} \left[
       2 \left( \mbox{Li}_2(-\rho x)+\mbox{Li}_2(-\rho y) \right) + \ln(\rho x) \ln(\rho y) 
       + \ln \left( \frac{y}{x} \right) \ln \left( \frac{1+\rho x}{1+\rho y} \right) + \frac{\pi^2}{3} \right] 
 \nonumber \\
 & & 
       + \frac{i \pi \theta(p_2^2)}{\sqrt{\Delta_3}} \ln \left( 
         \frac{\left(\delta_1+\sqrt{\Delta_3}\right) \left(\delta_3+\sqrt{\Delta_3}\right)}
              {\left(\delta_1-\sqrt{\Delta_3}\right) \left(\delta_3-\sqrt{\Delta_3}\right)} \right)
 + {\cal O}(\eps),
\eq
where
\bq
 x = \frac{p_1^2}{p_3^2}, \;\;\;\; y = \frac{p_2^2}{p_3^2}, \;\;\;\;
 \rho = \frac{2 p_3^2}{\delta_3+\sqrt{\Delta_3}}.
\eq
The step function $\theta(x)$ is defined as $\theta(x)=1$ for $x>0$ and $\theta(x)=0$ otherwise.

\subsection{Four-point functions}

For the four-point function we use the invariants
\bq
s = \left(p_1+p_2\right)^2,
 & &
t = \left(p_2+p_3\right)^2
\eq
together with the external masses $m_i^2=p_i^2$.
\\
The zero-mass box ($m_1^2=m_2^2=m_3^2=m_4^2=0$):
\bq
I^{0m}_4\left(s,t,\mu^2\right)  & = & 
 \frac{4}{\eps^2 s t} - \frac{2}{\eps s t} 
                      \left[ \ln\left(\frac{-s}{\mu^2}\right) + \ln\left(\frac{-t}{\mu^2}\right) \right]
 \nonumber \\
 & &
 + \frac{1}{s t} \left[ \ln^2\left(\frac{-s}{\mu^2}\right) + \ln^2\left(\frac{-t}{\mu^2}\right)
                       - \ln^2\left(\frac{-s}{-t}\right) - 8 \zeta_2
                 \right]
 + {\cal O}(\eps).
\eq
The one-mass box ($m_1^2=m_2^2=m_3^2=0$):
\bq
\lefteqn{
I^{1m}_4\left(s,t,m_4^2,\mu^2\right) = 
 \frac{2}{\eps^2 s t} - \frac{2}{\eps s t} 
                      \left[ \ln\left(\frac{-s}{\mu^2}\right) + \ln\left(\frac{-t}{\mu^2}\right) 
                             - \ln\left(\frac{-m_4^2}{\mu^2}\right) \right]
 + \frac{1}{s t} \left[ \ln^2\left(\frac{-s}{\mu^2}\right) 
 \right.
 } & &
 \nonumber \\
 & & \left.
 + \ln^2\left(\frac{-t}{\mu^2}\right)
                       - \ln^2\left(\frac{-m_4^2}{\mu^2}\right)
                       - \ln^2\left(\frac{-s}{-t}\right) 
                       - 2 \; \mbox{Li}_2\left(1- \frac{(-m_4^2)}{(-s)}\right)
                       - 2 \; \mbox{Li}_2\left(1- \frac{(-m_4^2)}{(-t)}\right)
 \right.
 \nonumber \\
 & & \left.
                       - 3 \zeta_2
                 \right]
 + {\cal O}(\eps).
\eq
The easy two-mass box ($m_1^2=m_3^2=0$):
\bq
\lefteqn{
I^{2me}_4\left(s,t,m_2^2,m_4^2,\mu^2\right) = 
 - \frac{2}{\eps \left( s t - m_2^2 m_4^2 \right)} 
                      \left[ \ln\left(\frac{-s}{\mu^2}\right) + \ln\left(\frac{-t}{\mu^2}\right) 
                             - \ln\left(\frac{-m_2^2}{\mu^2}\right) - \ln\left(\frac{-m_4^2}{\mu^2}\right) \right]
 } & &
 \nonumber \\
 & &
 + \frac{1}{s t - m_2^2 m_4^2} 
     \left[ \ln^2\left( \frac{-s}{\mu^2}\right) + \ln^2\left(\frac{-t}{\mu^2}\right)
                        - \ln^2\left(\frac{-m_2^2}{\mu^2}\right) - \ln^2\left(\frac{-m_4^2}{\mu^2}\right)
                       - \ln^2\left(\frac{-s}{-t}\right) 
 \right. \nonumber \\
 & & \left.
                       - 2 \; \mbox{Li}_2\left(1- \frac{(-m_2^2)}{(-s)}\right)
                       - 2 \; \mbox{Li}_2\left(1- \frac{(-m_2^2)}{(-t)}\right)
                       - 2 \; \mbox{Li}_2\left(1- \frac{(-m_4^2)}{(-s)}\right)
 \right. \nonumber \\
 & & \left.
                       - 2 \; \mbox{Li}_2\left(1- \frac{(-m_4^2)}{(-t)}\right)
                       + 2 \; \mbox{Li}_2\left(1- \frac{(-m_2^2)}{(-s)} \frac{(-m_4^2)}{(-t)}\right)
                 \right]
 + {\cal O}(\eps).
\eq
The hard two-mass box ($m_1^2=m_2^2=0$):
\bq
\lefteqn{
I^{2mh}_4\left(s,t,m_3^2,m_4^2,\mu^2\right) = 
 \frac{1}{\eps^2 s t} - \frac{1}{\eps s t} 
                      \left[ \ln\left(\frac{-s}{\mu^2}\right) + 2 \ln\left(\frac{-t}{\mu^2}\right) 
                             - \ln\left(\frac{-m_3^2}{\mu^2}\right) - \ln\left(\frac{-m_4^2}{\mu^2}\right) \right]
 } & &
 \nonumber \\
 & &
 + \frac{1}{s t} 
     \left[ \frac{3}{2} \ln^2\left( \frac{-s}{\mu^2}\right) + \ln^2\left(\frac{-t}{\mu^2}\right)
                        - \frac{1}{2} \ln^2\left(\frac{-m_3^2}{\mu^2}\right) 
                        - \frac{1}{2} \ln^2\left(\frac{-m_4^2}{\mu^2}\right)
                        - \ln^2\left(\frac{-s}{-t}\right) 
 \right. \nonumber \\
 & & \left.
                        - \ln\left(\frac{-s}{\mu^2}\right) \ln\left(\frac{-m_3^2}{\mu^2}\right) 
                        - \ln\left(\frac{-s}{\mu^2}\right) \ln\left(\frac{-m_4^2}{\mu^2}\right) 
                        + \ln\left(\frac{-m_3^2}{\mu^2}\right) \ln\left(\frac{-m_4^2}{\mu^2}\right) 
 \right. \nonumber \\
 & & \left.
                       - 2 \; \mbox{Li}_2\left(1- \frac{(-m_3^2)}{(-t)}\right)
                       - 2 \; \mbox{Li}_2\left(1- \frac{(-m_4^2)}{(-t)}\right)
                       - \frac{1}{2} \zeta_2
                 \right]
 + {\cal O}(\eps).
\eq
The three-mass box ($m_1^2=0$):
\bq
\lefteqn{
I^{3m}_4\left(s,t,m_2^2,m_3^2,m_4^2,\mu^2\right) = 
 - \frac{1}{\eps \left( s t - m_2^2 m_4^2 \right)} 
                      \left[ \ln\left(\frac{-s}{\mu^2}\right) + \ln\left(\frac{-t}{\mu^2}\right) 
                             - \ln\left(\frac{-m_2^2}{\mu^2}\right) - \ln\left(\frac{-m_4^2}{\mu^2}\right) \right]
 } & &
 \nonumber \\
 & &
 + \frac{1}{s t - m_2^2 m_4^2} 
     \left[   \frac{3}{2} \ln^2\left( \frac{-s}{\mu^2}\right) 
            + \frac{3}{2} \ln^2\left(\frac{-t}{\mu^2}\right)
            - \frac{1}{2} \ln^2\left(\frac{-m_2^2}{\mu^2}\right) 
            - \frac{1}{2} \ln^2\left(\frac{-m_4^2}{\mu^2}\right)
            - \ln^2\left(\frac{-s}{-t}\right) 
 \right. \nonumber \\
 & & \left.
            - \ln\left(\frac{-s}{\mu^2}\right) \ln\left(\frac{-m_3^2}{\mu^2}\right) 
            - \ln\left(\frac{-s}{\mu^2}\right) \ln\left(\frac{-m_4^2}{\mu^2}\right) 
            + \ln\left(\frac{-m_3^2}{\mu^2}\right) \ln\left(\frac{-m_4^2}{\mu^2}\right) 
 \right. \nonumber \\
 & & \left.
            - \ln\left(\frac{-t}{\mu^2}\right) \ln\left(\frac{-m_2^2}{\mu^2}\right) 
            - \ln\left(\frac{-t}{\mu^2}\right) \ln\left(\frac{-m_3^2}{\mu^2}\right) 
            + \ln\left(\frac{-m_2^2}{\mu^2}\right) \ln\left(\frac{-m_3^2}{\mu^2}\right) 
 \right. \nonumber \\
 & & \left.
            - 2 \; \mbox{Li}_2\left(1- \frac{(-m_2^2)}{(-s)}\right)
            - 2 \; \mbox{Li}_2\left(1- \frac{(-m_4^2)}{(-t)}\right)
            + 2 \; \mbox{Li}_2\left(1- \frac{(-m_2^2)}{(-s)} \frac{(-m_4^2)}{(-t)}\right)
                 \right]
 + {\cal O}(\eps).
\eq
The four-mass box:
\bq
I^{4m}_4\left(s,t,m_2^2,m_3^2,m_4^2,\mu^2\right) & = & I_3^{3m}(s t, m_1^2 m_3^2, m_2^2 m_4^2,\mu^2) + K(s, t, m_1^2, m_3^2, m_2^2, m_4^2),
\eq
where
\bq
\lefteqn{
K(s_1,t_1,s_2,t_2,s_3,t_3) = -\frac{2\pi i}{\lambda}
 \sum\limits_{i=1}^{3} \theta(-s_i) \theta(-t_i) 
} & & \nonumber \\
 & & \times
 \left[  
       \ln\left( \sum\limits_{j \neq i} s_j t_j - \left( s_i t_i - \lambda \right) (1+i 0) \right)
     - \ln\left( \sum\limits_{j \neq i} s_j t_j - \left( s_i t_i + \lambda \right) (1+i 0) \right)
 \right],
\eq
and
\bq
 \lambda & = & \sqrt{ \left( s_1 t_1 \right)^2 + \left( s_2 t_2 \right)^2 + \left( s_3 t_3 \right)^2
                      - 2 s_1 t_1 s_2 t_2 - 2 s_2 t_2 s_3 t_3 - 2 s_3 t_3 s_1 t_1 }.
\eq


\section{Analytic continuation}

In one-loop integrals the functions
\bq
 \ln\left(\frac{-s}{-t}\right), & &
 \mbox{Li}_2\left(1- \frac{(-s)}{(-t)} \right)
\eq
and generalizations thereof
occur. The analytic continuation is defined by giving all quantities a small
imaginary part, e.g.
\bq 
 s \rightarrow s + i 0.
\eq
Explicitly, the imaginary parts of the logarithm and the dilogarithm are given by
\bq
 \ln\left( \frac{-s}{-t} \right) & = & 
    \ln\left(\left|\frac{s}{t}\right|\right) -i \pi \left[ \theta(s) - \theta(t) \right],
 \nonumber \\
 \mbox{Li}_2\left(1 - \frac{(-s)}{(-t)} \right) & = & 
    \mbox{Re} \mbox{Li}_2\left(1 - \frac{s}{t} \right) 
    - i \theta\left(-\frac{s}{t}\right)
        \ln\left(1-\frac{s}{t}\right) 
        \mbox{Im} \ln \left(\frac{-s}{-t}\right).
\eq
This generalizes as follows:
\bq
 \ln\left( \frac{(-s_1)}{(-t_1)} \frac{(-s_2)}{(-t_2)} \right) & = & 
    \ln\left(\left|\frac{s_1 s_2}{t_1 t_2}\right|\right) 
    -i \pi \left[ \theta(s_1) + \theta(s_2) - \theta(t_1) - \theta(t_2) \right],
 \nonumber \\
 \mbox{Li}_2\left(1 - \frac{(-s_1)}{(-t_1)} \frac{(-s_2)}{(-t_2)} \right) & = & 
    \mbox{Re} \mbox{Li}_2\left(1 - \frac{s_1 s_2}{t_1 t_2} \right) 
    - i 
        \ln\left(1-\frac{(-s_1)}{(-t_1)} \frac{(-s_2)}{(-t_2)} \right) 
        \mbox{Im} \ln \left(\frac{(-s_1)}{(-t_1)} \frac{(-s_2)}{(-t_2)} \right), 
 \nonumber 
\eq
where
\bq
 \ln\left(1-\frac{(-s_1)}{(-t_1)} \frac{(-s_2)}{(-t_2)} \right)
 & = &
 \ln\left|1-\frac{s_1 s_2}{t_1 t_2} \right| 
 - \frac{1}{2} i \pi
   \left[ \theta(s_1) + \theta(s_2) - \theta(t_1) - \theta(t_2) \right]
   \theta\left( \frac{s_1 s_2}{t_1 t_2} - 1 \right).
 \nonumber
\eq


\section{Numerical evaluation of special functions}

The real part of the dilogarithm $\mbox{Li}_2(x)$ is numerically
evaluated as follows: Using the relations
\bq
 \mbox{Li}_2(x) & = & 
   -\mbox{Li}_2(1-x) + \frac{\pi^2}{6} -\ln(x) \ln(1-x),
 \nonumber \\
 \mbox{Li}_2(x) & = & 
   -\mbox{Li}_2\left(\frac{1}{x}\right) -\frac{\pi^2}{6}
   -\frac{1}{2} \left( \ln(-x) \right)^2,
\eq
the argument is shifted into the range $-1 \leq x \leq 1/2$. Then
\bq
 \mbox{Li}_2(x) & = & 
   \sum\limits_{i=0}^\infty \frac{B_i}{(i+1)!} z^{i+1} 
 \nonumber \\
 & = & B_0 z + \frac{B_1}{2} z^2 + \sum\limits_{n=1}^\infty \frac{B_{2 n}}{(2 n + 1)!} z^{2 n+ 1},
\eq
with $z = - \ln(1-x)$ and the $B_i$ are the Bernoulli numbers.
The Bernoulli numbers $B_i$ are defined through the generating function
\bq
 \frac{t}{e^t-1} & = & \sum\limits_{i=0}^\infty B_n \frac{t^n}{n!}.
\eq
It is also convenient to use the Clausen function $\mbox{Cl}_2(x)$ as an auxiliary function.
The Clausen function is given in terms of dilogarithms by
\bq
\label{defclausen}
 \mbox{Cl}_2(x) & = & \frac{1}{2i} \left[ \mbox{Li}_2\left(e^{ix}\right) - \mbox{Li}_2\left(e^{-ix}\right) \right].
\eq
Alternative definitions for the Clausen function are
\bq
 \mbox{Cl}_2(x) = \sum\limits_{n=1}^\infty \frac{\sin(n x)}{n^2} = - \int\limits_0^x dt \ln \left( \left|
                   2 \sin \left( \frac{t}{2} \right) \right| \right).
\eq
The Clausen function is evaluated numerically as follows:
Using the symmetry
\bq
 \mbox{Cl}_2(-x) & = & -\mbox{Cl}_2(x),
\eq
the periodicity
\bq
 \mbox{Cl}_2(x+2n \pi) & = & \mbox{Cl}_2(x),
\eq
and the duplication formula
\bq
 \mbox{Cl}_2(2x) & = & 2\mbox{Cl}_2(x)-2\mbox{Cl}_2(\pi-x)
\eq
the argument may be shifted into the range $ 0 \leq x \leq 2 \pi /3$.
Then
\bq
 \mbox{Cl}_2(x) & = & -x \ln(x) + x 
     + \sum\limits_{n=1}^\infty \frac{(-1)^{n+1}B_{2n}}{2 n (2 n + 1)!}x^{2n+1}.
\eq

\end{appendix}


\end{document}